\documentclass[acmsmall, nonacm]{acmart}
\acmNumber{HATRA} % CONF = POPL or ICFP or OOPSLA
\acmYear{2022}
\acmMonth{12}
\acmDOI{} % \acmDOI{10.1145/nnnnnnn.nnnnnnn}
\startPage{1}

\setcopyright{none}
\usepackage{booktabs}   %% For formal tables:
\usepackage{subcaption} %% For complex figures with subfigures/subcaptions
\usepackage{eulervm}
\usepackage{soul}
\usepackage{cleveref}
\usepackage{mathpartir}
\usepackage{tikz}
\begin{document}

%% Title information
\title{Holbert: Reading, Writing, Proving and Learning in the Browser}
%workshop this...

         %% [Short Title] is optional;
                                        %% when present, will be used in
                                        %% header instead of Full Title.
%\titlenote{with title note}             %% \titlenote is optional;
                                        %% can be repeated if necessary;
                                        %% contents suppressed with 'anonymous'
%\subtitle{Subtitle}                     %% \subtitle is optional
%\subtitlenote{with subtitle note}       %% \subtitlenote is optional;
                                        %% can be repeated if necessary;
                                        %% contents suppressed with 'anonymous'

%% Author information
%% Contents and number of authors suppressed with 'anonymous'.
%% Each author should be introduced by \author, followed by
%% \authornote (optional), \orcid (optional), \affiliation, and
%% \email.
%% An author may have multiple affiliations and/or emails; repeat the
%% appropriate command.
%% Many elements are not rendered, but should be provided for metadata
%% extraction tools.

%% Author with single affiliation.
\author{Liam O'Connor}
%\authornote{with author1 note}          %% \authornote is optional;
                                        %% can be repeated if necessary
\orcid{0000-0003-2765-4269}             %% \orcid is optional
\affiliation{
  \position{Lecturer}
  \department{School of Informatics}              %% \department is recommended
  \institution{University of Edinburgh}            %% \institution is required
%  \streetaddress{Street1 Address1}
  \city{Edinburgh}
%  \state{State1}
%  \postcode{Post-Code1}
  \country{Scotland} %% \country is recommended
}
\email{l.oconnor@ed.ac.uk}          %% \email is recommended

%% Author with two affiliations and emails.
\author{Rayhana Amjad}
%\authornote{with author2 note}          %% \authornote is optional;
                                        %% can be repeated if necessary
\orcid{0000-0002-3086-1720}             %% \orcid is optional
\affiliation{
  \position{PhD Student}
  \department{School of Informatics}             %% \department is recommended
  \institution{University of Edinburgh}           %% \institution is required
%  \streetaddress{Street2a Address2a}
  \city{Edinburgh}
%  \state{State2a}
%  \postcode{Post-Code2a}
  \country{Scotland}                   %% \country is recommended
}
\email{rayhana.yasmin.h@gmail.com}         %% \email is recommended

%% Abstract
%% Note: \begin{abstract}...\end{abstract} environment must come
%% before \maketitle command
\begin{abstract}
This paper presents Holbert a work-in-progress pedagogical proof assistant and online textbook platform, aimed at the educational use-case, specifically for the teaching of programming language theory. Holbert allows proof exercises and rule definitions to be embedded directly in an online textbook, where proofs and rules can be manipulated using a graphical interface. We give an overview of the logical foundations of Holbert, examples of its use, and give an update as to its current implementation status.
\end{abstract}

\newcommand{\liam}[1]{\hl{#1\textbf{ - Liam}}}
\newcommand{\rayhana}[1]{\hl{#1\textbf{ - Rayhana}}}

\newcommand{\holrule}[4]{\text{\tiny $#2$}\; \inferrule*[right={\tiny \(#1\)}]{#4}{#3}}
\newcommand{\holvar}[1]{\textcolor{green!40!black}{#1}}
\newcommand{\holbound}[1]{\textcolor{violet!90!white}{#1}}
\newcommand{\holelim}[1]{\textcolor{cyan!90!black}{#1}}
\newcommand{\holunif}[1]{\textcolor{gray!60!black}{#1}}
\newcommand{\holvdash}[0]{\mathbin{\textcolor{red!50!black}{\vdash}}}
\newcommand{\holcomma}[0]{\mathbin{\textcolor{red!50!black}{,}}}
\newcommand{\holreddish}[1]{\textcolor{red!50!black}{#1}}
\newcommand{\holassm}[1]{{\textcolor{blue!50!black}{\textsf{#1}}}}
\newcommand{\holleft}[0]{\textcolor{red!50!black}{(}}
\newcommand{\holright}[0]{\textcolor{red!50!black}{)}}
\newcommand{\holgoal}[0]{\begin{tikzpicture}[scale=0.4]
	\draw[ thick,blue] ([shift=(-30:0.3cm)]2,1) arc (-30:210:0.3cm) -- (2,0.55) -- ([shift=(-30:0.3cm)]2,1);
	\draw[,blue] (2,1) circle (0.13cm);
\end{tikzpicture}}
%% Keywords
%% comma separated list

%% \maketitle
%% Note: \maketitle command must come after title commands, author
%% commands, abstract environment, Computing Classification System
%% environment and commands, and keywords command.
\maketitle

\section{Introduction}

%ITP in classroom
Programming language theory is rife with valuable ideas that can be applied in the wider world of software, but it can also be challenging to teach; it requires a familiarity with mathematical proofs that students of computer science do not always possess. Teaching the ability to write proofs requires that students practice the skill with oversight to point out their errors and help them achieve correct proofs. The amount of feedback and interaction between teachers and students that this requires is not always practical, given large class sizes. To this end, \citet{pierce} advocates the use of proof assistants to more effectively communicate the foundations of programming language theory, as a proof assistant can similarly point out errors and verify the correctness of the proofs that a student writes.
 Early results from evaluation studies~\citep{evalstudy} show that students can profit strongly from the use of proof assistants in the classroom, compared to pen and paper proofs.\\
%%How has it been attempted?:
\Citet{pierce} uses the venerable proof assistant \citet{coq} in their classroom, based on the accompanying textbook, Software Foundations~\citep{software_foundations}. 
\citet{coqpl19} also use this book as the basis for a tool-assisted discrete mathematics course.
Inspired by Pierce's example, \citet{plfa} and \citet{nipkowklein} have also written similar textbooks for their favourite proof assistants. \citet{narboux} also uses Coq, but to teach mathematics, not programming language theory. \Citet{buzzard} does something similar with Lean. 

All of these proof assistants, however, are not primarily designed for a pedagogical purpose. They can be difficult software to use, with installing and executing the software proving an initial obstacle, particularly with large groups of students. The interfaces of these proof assistants typically involve unfamiliar syntax, notation and interaction idioms, and do not give simple feedback when a student provides an incorrect proof \citep{eastlund2007,bartzia2022proof}. Using such a proof assistant in a course runs the risk of shifting the emphasis away from the primary theoretical content of the course, and towards learning how to use the proof assistant. The fact that many people work through the above-mentioned textbooks in order to learn how to use a new proof assistant, rather than to learn the associated theory, is evidence of this phenomenon. Furthermore, the text-based interface of these provers means that students must copy example proofs from the textbook into their editors to examine a proof in their proof assistant, necessitating constant mental context-switching that adds unnecessary cognitive overhead.

 %\liam{blah blah something about how proof assistants are hard to use, difficult to install, and this makes them not good for teaching. Something about how insistence on soundness and rigorous foundations leads to complex core theories that take a lot of teaching time. Something about how students have to constantly context-switch between the textbook and the prover. Something about how the courses end up becoming more about the tool than the theory. It would be great if we could get some citations to back up some of these claims..}

Holbert is a new proof assistant that is designed to address these problems. It runs in the browser, so no installation is required. Rather than a text-based interface with complex syntax, Holbert presents rules, proofs and terms graphically, much as one would write them in a computer science paper. The interface is designed to fade into the background, and not require significant training to use. We provide several examples of its use in \Cref{sec:foundations}.

Crucially, Holbert is also an \emph{online textbook platform}. With Holbert, students do not have to switch between using their proof assistant and reading an accompanying textbook, copying code into their editors. Instead, students can simply interact with proof exercises directly \emph{in the textbook itself}. Online textbooks have the potential to be an exciting, interactive learning platform (and have already shown promising results in trials~\citep{edgcomb2015student}), but for the teaching of mathematics or theoretical computer science, this potential has not yet been fully realised. With Holbert, we aim to reach this potential, and unlock new and exciting ways to teach our favourite mathematical topics.

This paper is, itself, also available as a Holbert document, and is itself an example of an interactive Holbert document\footnote{Some conversion is required to fit the required ACM paper format, however.}. 

To view the interactive version of this paper, see \texttt{\url{http://liamoc.net/hatra-2022}}.

%% - Successes
%% - Obstacles

\section{A Tour of Holbert}
\label{sec:foundations}

The foundation of Holbert is an \emph{untyped} variant of the original \emph{higher order logic} of \citet{church}, which uses $\lambda$-calculus to represent logical terms. Higher order logic is also the basis of proof assistants such as Isabelle/HOL~\citep{isabelle}, but these proof assistants use typed formalisms to escape the paradoxes that are present in the untyped theory.\footnote{Consider that the untyped $\lambda$-term $(\mathcal{Y}\ \neg)$, where $\mathcal{Y}$ is a fixed-point combinator, $\beta$-reduces to its own negation.} The main reason for the elision of a type system is pedagogical. Holbert is intended as a vehicle for teaching programming language theory, including type theory, and requiring type theory knowledge \emph{a priori} to effectively use Holbert would be an irritating circularity. The lack of a type system also generally reduces the number of concepts required to learn Holbert, and enables our definitions to more closely resemble the untyped, informal definitions used in conventional computer science notation. Of course, this does mean that Holbert is technically unsound, but the theorems proven in Holbert are not intended for, say, verifying an operating system kernel~\citep{Klein:2009}---there is no requirement that Holbert theorems be trustworthy. To paraphrase the popular phrase, ``\emph{Doctor, I can prove $\bot$ when I do this!}''---then don't do that!

\subsection{Defining Rules}

Logical statements, such as rules, theorems and axioms, are represented in Holbert using the format of natural deduction~\citep{gentzen} inference rules, specifically \emph{hereditary Harrop formulae}~\citep{harrop}. A hereditary Harrop formula is of the format {\small $\text{\tiny\(\overline{\holvar{A}.}\)}\  \overline{H} \holvdash C $}, stating that for all substitutions of the metavariables {\small $\overline{\holvar{A}}$}, the conclusion {\small $C$} holds if the premises {\small $\overline{H}$}, which are themselves hereditary Harrop formulae, hold. Here $\overline{\text{overlines}}$ represent a sequence of zero or more. This structure allows for all of the usual natural deduction rules of intuitionistic propositional logic to be encoded in a natural way.

\vspace{0.4em}
\noindent\textbf{Axioms.}
{\small \begin{gather*}
\holrule{\land{}I}{\holvar{A.}\holvar{B.} }{\holvar{A} \land \holvar{B}}{\holvar{A}\quad\holvar{B}}\quad
\holrule{\land{}E_1}{\holvar{A.}\holvar{B.} }{\holvar{A}}{\holvar{A}\land\holvar{B}}\quad
\holrule{\land{}E_2}{\holvar{A.}\holvar{B.} }{\holvar{B}}{\holvar{A}\land\holvar{B}}\quad
\holrule{\rightarrow\!\!I}{\holvar{A.}\holvar{B.}}{\holvar{A} \rightarrow \holvar{B}}{\holvar{A} \holvdash \holvar{B}}\quad
\holrule{\rightarrow\!\!E}{\holvar{A.}\holvar{B.}}{\holvar{B}}{\holvar{A} \rightarrow \holvar{B}\quad\holvar{A}}\\[0.3em]
\holrule{\lor{}I_1}{\holvar{A.}\holvar{B.} }{\holvar{A}\lor\holvar{B}}{\holvar{A}}\;\;
\holrule{\lor{}I_2}{\holvar{A.}\holvar{B.} }{\holvar{A}\lor\holvar{B}}{\holvar{B}}\;\;
\holrule{\lor{}E}{\holvar{A.}\holvar{B.}\holvar{C.} }{\holvar{C}}{\holvar{A}\lor\holvar{B}\quad \holvar{A}\holvdash\holvar{C}\quad \holvar{B}\holvdash\holvar{C}}\;\;
\holrule{\neg{}I}{\holvar{P.}}{\neg\ \holvar{P}}{\text{\tiny$\holvar{X.}$}\ \holvar{P} \holvdash \holvar{X}}\;\;
\holrule{\neg{}E}{\holvar{P.}\holvar{X.}}{\holvar{X}}{\neg\ \holvar{P}\quad \holvar{P}}
\end{gather*}}
By default, Holbert renders rules in the \emph{hybrid} format, where a vinculum is used for entailment on the top level, but the entailment sign ($\holvdash$) is used for hypothetical derivations. Holbert also supports \emph{linear} style, where rules are always rendered horizontally, and \emph{vertical} style, which is exactly Gentzen's original notation, using vincula for top-level entailments and a vertical layout with dots for hypothetical derivations.

Any term in a rule definition, such as those above, can be edited in Holbert simply by clicking on it. Infix operators, such as {\small $\land$}, are actually applied prefix, so, for example, the conclusion of the rule {\small $\land{}I$} is actually written as \texttt{\_/\textbackslash\_ A B}. Allowing terms to be edited with infix syntax is future work.

\subsection{Proofs}

Proofs are represented as hierarchical trees of goals, similar to \citet{gentzen}. Each stage of the proof may introduce new bound metavariables, and new assumptions, which are given numeric names in Holbert. Below is an incomplete proof of the commutativity of conjunction, where a \emph{goal tag} $\holgoal$ is left at the unsolved goal. Clicking on the goal tag will open a goal summary on the right-hand panel of the Holbert interface, where the current goal, assumptions and applicable rules are displayed. 

\vspace{0.4em}
\noindent\textbf{Theorem.}\\
{\small 
 	$\holrule{\land{}\mathsf{comm}}{\holvar{A. B.}}{\holleft\holvar{A} \land \holvar{B}\holright \rightarrow \holleft\holvar{B} \land \holvar{A}\holright}{$ $}$}\\[0.2em]
\textit{Proof}.\\
{\small 
 	$\holrule{\rightarrow{}\!\!I}{\holvar{A. B.}}{\holleft\holvar{A} \land \holvar{B}\holright \rightarrow \holleft\holvar{B} \land \holvar{A}\holright}{\holrule{\land{}I}{}{\holassm{0:}\ \holvar{A}\land\holvar{B} \holvdash \holvar{B} \land \holvar{A}}
 	      {\holrule{\land{}E_2}{}{\holvar{B}}{\holrule{\holassm{0}}{}{\holvar{A}\land\holvar{B}}{ }}\quad \holrule{?}{}{\holvar{A}}{\holgoal}}}$}\\[0.5em]
If an applicable rule is clicked, the rule is applied to the current goal as an introduction rule in the ``backwards reasoning'' commonly used in natural deduction. By default, only introduction rules are displayed, as all elimination rules would be applicable in this sense to all goals. While Holbert has special features for elimination rules, outlined in \Cref{sec:elim}, elimination rules may also be applied as introduction rules by toggling a checkbox in the goal display, to show rules that do not fit the introduction rule format.

When applying a rule {\small $\text{\tiny\(\overline{\holvar{A}.}\)}\ \overline{H}\ \holvdash\ C $} to a goal by backwards reasoning, we first replace all metavariables {\small $\overline{\holvar{A}}$} in the rule with fresh, global \emph{unification variables} (also called \emph{schematic variables}). Then, we attempt to find a substitution to unification variables that unifies the goal with the conclusion {\small $C$}. If such a substitution can be found, new subgoals are added for each premise {\small $\overline{H}$} and the substitution is applied to all unification variables occurring in the proof.

\subsection{Higher Order Unification}

Because our logic terms are $\lambda$-terms, the first order unification algorithm of \citet{robinson} is insufficient. Worse, the problem of unification modulo $\alpha\beta\eta$-equivalence is, in general, undecidable. \Citet{huet} provides a semi-decision procedure, but this remains an unwieldy solution. Instead, and similarly to many other proof assistants, Holbert instead makes use of the \emph{pattern unification} technique of \citet{miller}, specifically derived from the implementation of \citet{nipkow}. This technique gives most general unifiers for terms that fall within the \emph{pattern subset}, that is, terms where unification variables may only be applied to a list of distinct object variables. This restriction is not too onerous for the application of introduction rules, but some elimination and induction rules require some workarounds, described in \Cref{sec:elim,sec:induct}.

Because unification variables are (proof-)global in scope, any substitution for a unification variable cannot directly mention the bound metavariables in scope for the goal. Thus, if we have a goal to prove, say, {\small $\holvar{A}\land\holvar{A} $} for all {\small $\holvar{A}$} given an assumption {\small $\holassm{0:}\ \holvar{A}$}, and attempt to apply the rule {\small $\land{}I$}, instantiating the rule {\small $\land{}I$} with naked unification variables would not work, because the conclusion {\small $\holunif{?_1} \land \holunif{?_2}$} would not be unifiable with {\small $\holvar{A}\land\holvar{A}$}, as the assignments to the unification variables {\small $\holunif{?_1}$} and {\small $\holunif{?_2}$} cannot contain the bound metavariable {\small $\holvar{A}$}.  To address this problem, we instead \emph{apply all bound variables in scope for the goal} to the unification variables when we instantiate the rule. In our example, this means that we must now unify {\small $\holvar{A}\land\holvar{A} $} with {\small $\holleft \holunif{?_1}\ \holvar{A}\holright \land \holleft \holunif{?_2}\ \holvar{A}\holright$}, which can easily be solved by assigning the identity function to both unification variables.

\subsection{Higher Order Abstract Syntax}
We can make use of Holbert's support for $\lambda$-abstractions to write rules for quantifiers using $\lambda$-abstraction to handle variable binding. To reduce notational clutter, $\lambda$-abstractions in Holbert are written without the $\lambda$, so the identity function would just be written as \texttt{(x. x)}. 

\vspace{0.4em}
\noindent\textbf{Axioms.}
{\small \begin{gather*}
\holrule{\forall{}I}{\holvar{P.} }{\forall\ \holleft\holbound{a.}\ \holvar{P}\ \holbound{a}\holright}{\text{\tiny$\holvar{x.}$}\ \holvar{P}\ \holvar{x}}\quad
\holrule{\forall{}E}{\holvar{P.x.} }{\holvar{P}\ \holvar{x}}{\forall\ \holleft\holbound{a.}\ \holvar{P}\ \holbound{a}\holright}\quad
\holrule{\exists{}I}{\holvar{P.x.} }{\exists\ \holleft\holbound{a.}\ \holvar{P}\ \holbound{a}\holright}{\holvar{P}\ \holvar{x}}\quad
\holrule{\exists{}E}{\holvar{P.Q.} }{\holvar{Q}}{\exists\ \holleft\holbound{a.}\ \holvar{P}\ \holbound{a}\holright\quad\text{\tiny$\holvar{x.}$}\ \holvar{P}\ \holvar{x}\holvdash \holvar{Q}}
\end{gather*}}
This technique of \emph{higher order abstract syntax} re-uses the existing binding mechanisms in Holbert to provide very natural looking rules for logical quantifiers. The same technique can also be used to represent variables and binding in programming language syntax, enabling programming languages to be formalised without the pedagogical overhead induced by term representations with name strings and substitutions, or \citet{debruijn} indices.

\subsection{Elimination Rules}
\label{sec:elim}
The elimination rule {\small $\forall{}E$}, presented in the previous section, is somewhat problematic as, after instantiation, its conclusion consists of a unification variable (from {\small $\holvar{P}$}) applied to another unification variable (from {\small $\holvar{x}$}), which is outside the pattern subset. In practice, if we were to apply {\small $\forall{}E$} anyway, using backwards reasoning as with an introduction rule, to any goal {\small $G$}, our pattern unification would instantiate {\small $\holvar{P}$} to a function that \emph{ignores} its argument and simply returns {\small $G$}, meaning that {\small $\holvar{x}$} and the quantified variable {\small $\holbound{a}$} are completely ignored. This is obviously not desirable. 

But, applying elimination rules as introduction rules is, in any case, undesirable: While Gentzen's original presentation of natural deduction uses elimination rules in this way, this presents several usability problems, in addition to the aforementioned unification issue. Chiefly, as mentioned previously, the conclusion of any elimination rule can be unified with \emph{any} goal. Thus, \emph{all} elimination rules are applicable, in the sense of backwards reasoning, to a given goal, which presents an unwieldy list of candidate rules to the user. 

   In the process of constructing a proof, we think of applying elimination rules as \emph{forward reasoning}, not backwards. Almost always, an elimination rule is thought to be applied to an assumption in the current context, \emph{not} to the current goal. 
Thus, Holbert features special features for applying elimination rules. Instead of clicking a rule to apply backwards to the current goal, the user clicks an \emph{assumption} on which to apply an elimination rule, and from there applicable elimination rules are displayed and may be applied to the goal.

Specifically, when applying an elimination rule {\small $\text{\tiny\(\overline{\holvar{A}.}\)}\ P_0, \overline{H}\, \holvdash C $} to an assumption {\small $S$} for a goal {\small $G$}, we once again replace metavariables {\small $\overline{\holvar{A}}$} in the rule with fresh, global unification variables (with all bound variables in scope for the goal applied). Then, rather than unify {\small $C$} with our goal {\small $G$}, we first try to unify the first premise of the rule {\small $P_0$} with the assumption {\small $S$}, and only after this unification and substitution do we attempt to unify the conclusion of the rule with our goal. Note that, in the case of {\small $\forall{}E$}, this will avoid the non-pattern cases encountered previously. The first (and only) premise of the rule, {\small $\forall\ \holleft\holbound{a.}\ \holvar{P}\ \holbound{a}\holright$}, only contains one unification variable after instantiation. Thus the first unification will, falling within the pattern subset, find the correct assignment to instantiate $\holvar{P}$, thus eliminating the non-pattern scenario previously encountered when unifying {\small $C$} with {\small $G$}.

When an elimination rule is applied in this way, the first subgoal from the applied rule is omitted in the resulting tree, as it is trivially discharged by the assumption. Instead, the numeric name of the assumption is superscripted to the name of the applied rule. As an example, here we prove one direction of the de Morgan identity for quantifiers:

\vspace{0.4em}
\noindent\textbf{Theorem.}\nopagebreak\\[0.2em]
{\small 
 	$\holrule{\neg\forall\exists}{\holvar{P.}}{\neg\ \holleft\forall\ \holleft\holbound{a.}\ \holvar{P}\ \holbound{a}\holright\holright}{\exists\ \holleft\holbound{x.}\ \neg\ \holleft\holvar{P}\ \holbound{x}\holright\holright}$}\\[0.2em]
\textit{Proof}.\\
{\small 
 	$\holrule{\neg{}I}
 	         {\holvar{P.}}
 	         {\holassm{0:}\ \exists\ \holleft\holbound{x.}\ \neg\ \holleft\holvar{P}\ \holbound{x}\holright\holright \holvdash \neg\ \holleft\forall\ \holleft\holbound{a.}\ \holvar{P}\ \holbound{a}\holright\holright}
 	         {\holrule{\holelim{\exists{}E}^\holassm{0}}
 	                  {\holvar{F.}}
 	                  {\holassm{1:}\ \forall\ \holleft\holbound{a.}\ \holvar{P}\ \holbound{a}\holright \holvdash \holvar{F}}
 	                  {\holrule{\holelim{\neg{}E}^\holassm{2}}
 	                           {\holvar{x.}}
 	                           {\holassm{2:}\ \neg\ \holleft\holvar{P}\ \holvar{x} \holright\holvdash \holvar{F}}
 	                           {\holrule{\holelim{\forall{}E}^\holassm{1}}
 	                                    {}
 	                                    {\holvar{P}\ \holvar{x}}
 	                                    { }}}}$}
\subsection{Induction}
\label{sec:induct}
As a convenience feature, Holbert also includes support for inductive judgements. The user specifies only the introduction rules, and an induction principle and cases rule for inversion are synthesised automatically. To save space, we will simply define the natural numbers here, but these definitions may be significantly more complex. For example, Holbert can synthesise principles for simultaneous induction given mutually inductive introduction rules.

\vspace{0.4em}
\noindent\textbf{Inductive Definition.}\\[0.2em]
{\small 
 	$\holrule{\mathsf{zero}}{}{0\ \mathbb{N}}{ } \quad  	\holrule{\mathsf{suc}}{\holvar{n.}}{\holleft S\ \holvar{n}\holright\ \mathbb{N}}{\holvar{n}\ \mathbb{N}} $}\\[0.2em]
\textit{Derived Rules}\\[0.4em]
{\small
 	$\holrule{\mathbf{cases}(\text{\textunderscore}\mathbb{N})}{\holvar{P.x.}}{\holvar{P}}{\holvar{x}\ \mathbb{N}\quad \holvar{x} = 0 \holvdash \holvar{P} \quad \text{\tiny \(\holvar{n.}\)}\ \holvar{x} = \holleft S\ \holvar{n} \holright \holvdash \holvar{P}} \quad$}{\small
 	$\holrule{\mathbf{induction}(\text{\textunderscore}\mathbb{N})}{\holvar{P.x.}}{\holvar{P}\ \holvar{x}}{\holvar{x}\ \mathbb{N}\quad \holvar{P}\ 0 \quad \text{\tiny \(\holvar{n.}\)}\ \holvar{P}\ \holvar{n} \holcomma \holvar{n}\ \mathbb{N}  \holvdash \holvar{P}\ \holleft S\ \holvar{n} \holright} $}\\[0.2em]
Both of these derived rules can be applied as an elimination rule. The induction rule, however, exposes another scenario where our unification falls outside the pattern subset, and workarounds are required. Consider the case where we must prove some goal {\small $G\ \holvar{k}$} for all {\small$\holvar{k}$}, with the assumption that {\small$\holvar{k}\ \mathbb{N}$}. 
    We instantiate the rule with fresh unification variables, replacing {\small$\holvar{P}$} with {\small $\holleft\holunif{?_0}\ \holvar{k} \holright$} and {\small$\holvar{x}$} with {\small $\holleft\holunif{?_1}\ \holvar{k} \holright$}. Then we unify the first premise of our instantiated rule {\small$\holleft\holunif{?_1}\ \holvar{k} \holright\ \mathbb{N}$} with the assumption {\small$\holvar{k}\ \mathbb{N}$}, which resolves {\small$\holunif{?_1}$} to the identity function. We must now unify our goal {\small $G\ \holvar{k}$} with the conclusion of our rule, which is now {\small $\holleft\holunif{?_0}\ \holvar{k} \holright\ \holvar{k}$}. This falls outside the pattern subset, as {\small$\holvar{k}$} occurs twice in the arguments to a unification variable\footnote{Our unification algorithm once again picks the wrong solution here, choosing the first argument rather than the second.}. 
    
    To avoid this problem, we adjust our strategy for applying elimination rules slightly. Instead of instantiating all variables in the rule at once, we instantiate variables in two phases. Firstly, we instantiate all those variables that occur in the first premise of the rule. These are instantiated with fresh unification variables that are applied to all bound metavariables in the scope of the goal, as before. Then we unify the first premise of the instantiated rule with our assumption. After applying the resulting substitution to the rule, we instantiate the remaining variables in the rule with fresh unification variables, but now they are only applied to those bound variables in the scope of the goal which do not already occur in the substituted rule. Then we unify the goal and the conclusion as before. This avoids the problematic term {\small $\holleft\holunif{?_0}\ \holvar{k} \holright\ \holvar{k}$} we saw earlier, because {\small$\holvar{k}$} will only occur once in the second unification.
\subsection{Prose-style proofs}

While induction and proofs by cases can be presented as a natural deduction tree, this is not typically how such proofs are presented. The tree quickly becomes unwieldy and wide. Instead, Holbert allows proofs to be presented in \emph{prose-style}, which tends to grow vertically rather than horizontally, and allows the user to write expository text for each case. As an example, we can prove that any non-zero natural number must be the successor of some other natural number.

\vspace{0.4em}
\noindent\textbf{Theorem.}\nopagebreak\\[0.2em]
{\small 
 	$\holrule{\mathsf{pred}}{\holvar{n.}}{\exists\ \holleft\holbound{k}.\ \holvar{n} = \holleft S\ \holbound{k}\holright\holright}{\holvar{n}\ \mathbb{N}\quad \neg\ \holleft\holvar{n} = 0\holright}$}\\[0.2em]
\textit{Proof}.\\
\textit{Given} {\small $\holvar{n.}$} \textit{where:}
\begin{itemize}
\item {\small $\holassm{0:}\ \holvar{n}\ \mathbb{N}$	}
\item {\small $\holassm{1:}\ \neg\ \holleft\holvar{n} = 0\holright$	}
\end{itemize}
Show {\small $\exists\ \holleft\holbound{k}.\ \holvar{n} = \holleft S\ \holbound{k} \holright \holright$}\\
by {\small $\holelim{\mathbf{cases}(\text{\textunderscore}\mathbb{N})}^\holassm{0}$}:
\begin{itemize}
\item \textit{Assuming}\ {\small $\holassm{2:}\ \holvar{n} = 0$} \\
      We can see that $\holassm{2}$ contradicts $\holassm{1}$, so we can discharge our goal via \emph{ex falso quodlibet}:\\[0.2em]
      {\small ${}\quad\exists\  \holleft\holbound{k}.\ \holvar{n} = \holleft S\ \holbound{k} \holright \holright$}\\
      by:\\
      {\small $\holrule{\holelim{\neg{}E}^\holassm{1}}{}{\exists\  \holleft\holbound{k}.\ \holvar{n} = \holleft S\ \holbound{k} \holright \holright}{\holrule{\holassm{\scriptsize 0}}{}{\holvar{n} = 0}{ }}$}
\item \textit{Given} {\small $\holvar{k.}$} \textit{where}:
\begin{itemize}
\item {\small $\holassm{2:}\ \holvar{n} = \holleft S\ \holvar{k}\holright$	}
\item {\small $\holassm{3:}\ \holvar{k}\ \mathbb{N}$	}
\end{itemize}
In this case, assumption {\small $\holassm{2}$} trivially gives a witness for our goal:\\[0.2em]
      {\small ${}\quad\exists\  \holleft\holbound{k}.\ \holvar{n} = \holleft S\ \holbound{k} \holright \holright$}\\
      by:\\
      {\small $\holrule{\exists{}I}{}{\exists\  \holleft\holbound{k}.\ \holvar{n} = \holleft S\ \holbound{k} \holright \holright}{\holrule{\holassm{\scriptsize 2}}{}{\holvar{n} = \holleft S\ \holvar{k} \holright}{ }}$}\\
\end{itemize}
In this case, only the top level application of the cases rule is presented in prose-style, with each case still presented in tree-style. Holbert allows the reader to decide which style to use, and can switch between them seamlessly.

\subsection{Equality}
As can be seen in the derived cases rules, Holbert has a built-in notion of equality. It also supports rewriting a goal by such equalities. Of course, it is possible to encode such rewriting as an elimination rule, like:
{\small $$\holrule{\mathsf{subst}}{\holvar{P.x.y.}}{\holvar{P}\ \holvar{y}}{\holvar{x} = \holvar{y} \quad \holvar{P}\ \holvar{x}}$$}
But, as our elimination rules are typically applied to an assumption, this would only be practically useful if the equality was one of the assumptions inside our goal context. Thus, we instead provide a bespoke rewriting function. Clicking a button in the goal view displays available equalities by which to rewrite. To determine if an equality is applicable, Holbert searches through the goal for a subterm that unifies with the left (or right, if the rewriting direction is reversed) hand side of the equality. To demonstrate this feature, we will first axiomatise some equations to define addition of natural numbers.

\vspace{0.4em}
\noindent\textbf{Axioms.}\\[0.1em]
{\small $
\holrule{+B}{\holvar{n.}}{\holleft 0 + \holvar{n}\holright = \holvar{n}}{ }\quad
\holrule{+I}{\holvar{m.}\holvar{n.} }{\holleft\holleft S\ \holvar{m}\holright +\holvar{n}\holright = \holleft S\ \holleft\holvar{m}+\holvar{n}\holright\holright}{ }$}\\[0.3em]
We can then inductively prove the right identity of addition, making use of rewriting in the inductive case:

\vspace{0.4em}
\noindent\textbf{Theorem.}\nopagebreak\\[0.2em]
{\small 
 	$\holrule{\mathsf{pred}}{\holvar{n.}}{ \holleft n + \holvar{0}\holright = \holvar{n}}{\holvar{n}\ \mathbb{N}}$}\\[0.2em]
\textit{Proof}.\\
\textit{Given} {\small $\holvar{n.}$} \textit{where} {\small $\holassm{0:}\ \holvar{n}\ \mathbb{N}$	}\\
We shall show\\[0.2em]
{\small ${}\quad \holleft n + \holvar{0}\holright = \holvar{n}$}\\[0.2em]
by {\small $\holelim{\mathbf{induction}(\text{\textunderscore}\mathbb{N})}^\holassm{0}$}:
\begin{itemize}
\item For the base case, we can simply show\\[0.2em]
      {\small ${}\quad\holleft0 + 0\holright = 0$}\\[0.2em]
      by
      {\small $+B$}.\\[-0.8em]
\item \textit{Given} {\small $\holvar{k.}$} \textit{where}:
\begin{itemize}
\item {\small $\holassm{2:}\ \holleft\holvar{k} + 0\holright = \holvar{k}$	}
\item {\small $\holassm{3:}\ \holvar{k}\ \mathbb{N}$	}
\end{itemize}
This inductive case is shown by rewriting\\[0.2em]
{\small ${}\quad \holleft \holleft S\ \holvar{k}\holright + \holvar{0}\holright = \holleft S\ \holvar{k}\holright$}\\[0.2em]
      by:\\
      {\small $\holrule{\holreddish{+I^\rightarrow}}{}{\holleft \holleft S\ \holvar{k}\holright + \holvar{0}\holright = \holleft S\ \holvar{k}\holright}{\holrule{\holreddish{\holassm{\scriptsize 2}^\rightarrow}}{}{\holleft S\ \holleft\holvar{k} + \holvar{0}\holright\holright = \holleft S\ \holvar{k}\holright}{ 
          \holrule{\mathbf{refl}}{}{ \holleft S\ \holvar{k}\holright  = \holleft S\ \holvar{k}\holright}{ }}}$}
\end{itemize}
The arrow superscripts on the names of the rules applied indicate the direction of rewriting, and {\small $\mathbf{refl}$} is the built-in axiom of reflexivity. 

In proofs of equalities such as the above, it is common to present proofs as a sequence of equalities, in a \emph{calculational} or \emph{equational} style. Currently, this is not supported by Holbert, but work is underway to implement it, for both equalities and, more generally, for any preorder.
\section{Related Work}

Holbert is not the only graphical proof assistant, nor the only proof assistant designed for education. Logitext~\citep{logitext} runs in the browser and is used to make an interactive tutorial for the sequent calculus~\citep{interactivesequent}. Like Holbert, it features a graphical interface for constructing Gentzen trees. But, unlike Holbert, it is limited to the connectives and quantifiers of first-order logic, and does not allow students nor teachers to define their own rules or definitions.
Jape~\citep{jape} is a now-defunct graphical prover written in Java that allows students to explore proofs using unification in a pre-encoded logic. \citet{edukera} and \citet{deaduction} are both similar graphical proof interfaces, for Coq and Lean respectively. As \citet{bartzia2022proof} note, all of these tools are not as general-purpose as Holbert, as they focus on \emph{proof} exercises and do not consider \emph{formalisation} exercises. They do not allow student users to create new rules or definitions\footnote{According to \citet{bartzia2022proof}, Edukera further restricts teachers to simply composing exercise sheets out of pre-defined exercises.}. This prevents the setting of exercises where a student must formalise in the proof assistant a theory described to them in writing, for example.

Alfa~\citep{alfa}, an earlier incarnation of Agda~\citep{agda}, is a proof assistant based on Martin-L\"of type theory with a structural editor and visualisation of proofs based on Gentzen trees. While Alfa is higher-order, graphical, and general purpose, it is no longer maintained, and unlike Holbert it is not suitable as a platform for a textbook.

SASyLF~\citep{sasylf} is an educational proof assistant intended for programming language theory. It is foundationally based on Twelf~\citep{twelf}, and therefore supports higher order abstract syntax as Holbert does. Unlike Holbert, it uses a more conventional textual interface, but with a syntax that tries to approximate conventional computer science notation.  

ORC$^2$A~\citep{orc2a} is a proof assistant that aims to make proof construction approachable to computer science students already skilled at programming, specifically applied to verification of computer programs. It features a text-based explicit proof language similar to Isabelle's Isar language.

Unlike Holbert, none of the above-mentioned proof assistants are suitable for use as an interactive textbook platform. Lurch~\citep{lurch}, on the other hand, is primarily intended as a mathematical word processor, but includes some proof checking capabilities via its OpenMath backend. Unlike Holbert, this proof checking capability is more of a post-hoc sanity check\footnote{The authors describe it as a spell-checker for mathematics.}, and it does not assist the user in writing the proofs with an interactive interface, as Holbert does. 

\section{Conclusions}

Holbert is still a work in progress, but it can already be used as an educational aid. It has been used in classes on logic TU Dresden, and \citet{zabr} has successfully formalised the type theory of Martin-L\"of in Holbert and carried out some non-trivial proofs.

In addition to the calculational proofs mentioned above, we also intend to add support for data-type style inductive definitions, which will, in addition to induction and cases rules, also generate disjointness and injectivity axioms to enable the convenient definition of inductive data types. This will in turn enable implementation of function definitions, which will bring Holbert roughly in line other HOL-based provers in terms of fundamental proof assistant features. There is also tremendous scope for domain-specific extensions to Holbert. Constructing category-theoretic proofs using commutative or string diagrams, doing geometry proofs visually, or annotating a computer program with Hoare logic assertions could be done as naturally as on paper. We are excited to explore these possibilities.

Holbert is an open-source project, implemented in Haskell and compiled to JavaScript using \citet{ghcjs}. We encourage contributions from anyone who is interested, and new features and capabilities are being added all the time. 
$$\text{https://github.com/liamoc/holbert}$$

%% Acknowledgments
\begin{acks}                            %% acks 
Many thanks to all contributors to the Holbert project, especially Chris Perceval-Maxwell and Yueyang Tang.
\end{acks}
%% Bibliography
\bibliography{cites}


\begin{thebibliography}{35}

%%% the Web address.

\ifx \showCODEN    \undefined \def \showCODEN     #1{\unskip}     \fi
\ifx \showDOI      \undefined \def \showDOI       #1{#1}\fi
\ifx \showISBNx    \undefined \def \showISBNx     #1{\unskip}     \fi
\ifx \showISBNxiii \undefined \def \showISBNxiii  #1{\unskip}     \fi
\ifx \showISSN     \undefined \def \showISSN      #1{\unskip}     \fi
\ifx \showLCCN     \undefined \def \showLCCN      #1{\unskip}     \fi
\ifx \shownote     \undefined \def \shownote      #1{#1}          \fi
\ifx \showarticletitle \undefined \def \showarticletitle #1{#1}   \fi
\ifx \showURL      \undefined \def \showURL       {\relax}        \fi
% invisible to TeX
\providecommand\bibfield[2]{#2}
\providecommand\bibinfo[2]{#2}
\providecommand\natexlab[1]{#1}
\providecommand\showeprint[2][]{arXiv:#2}

\bibitem[Aldrich et~al\mbox{.}(2008)]%
        {sasylf}
\bibfield{author}{\bibinfo{person}{Jonathan Aldrich},
  \bibinfo{person}{Robert~J. Simmons}, {and} \bibinfo{person}{Key Shin}.}
  \bibinfo{year}{2008}\natexlab{}.
\newblock \showarticletitle{SASyLF: An Educational Proof Assistant for Language
  Theory}. In \bibinfo{booktitle}{\emph{Proceedings of the 2008 International
  Workshop on Functional and Declarative Programming in Education}} (Victoria,
  BC, Canada) \emph{(\bibinfo{series}{FDPE '08})}.
  \bibinfo{publisher}{Association for Computing Machinery},
  \bibinfo{address}{New York, NY, USA}, \bibinfo{pages}{31–40}.
\newblock
\showISBNx{9781605580685}
\urldef\tempurl%
\url{https://doi.org/10.1145/1411260.1411266}
\showDOI{\tempurl}


\bibitem[Bartzia et~al\mbox{.}(2022)]%
        {bartzia2022proof}
\bibfield{author}{\bibinfo{person}{Evmorfia Bartzia}, \bibinfo{person}{Antoine
  Meyer}, {and} \bibinfo{person}{Julien Narboux}.}
  \bibinfo{year}{2022}\natexlab{}.
\newblock \showarticletitle{Proof assistants for undergraduate mathematics and
  computer science education: elements of a priori analysis}.
\newblock  (\bibinfo{year}{2022}).
\newblock


\bibitem[Bornat and Sufrin(1999)]%
        {jape}
\bibfield{author}{\bibinfo{person}{Richard Bornat} {and}
  \bibinfo{person}{Bernard Sufrin}.} \bibinfo{year}{1999}\natexlab{}.
\newblock \showarticletitle{{Animating Formal Proof at the Surface: The Jape
  Proof Calculator}}.
\newblock \bibinfo{journal}{\emph{Comput. J.}} \bibinfo{volume}{42},
  \bibinfo{number}{3} (\bibinfo{date}{01} \bibinfo{year}{1999}),
  \bibinfo{pages}{177--192}.
\newblock
\showISSN{0010-4620}
\urldef\tempurl%
\url{https://doi.org/10.1093/comjnl/42.3.177}
\showDOI{\tempurl}
\showeprint{https://academic.oup.com/comjnl/article-pdf/42/3/177/962523/420177.pdf}


\bibitem[Buzzard(2022)]%
        {buzzard}
\bibfield{author}{\bibinfo{person}{Kevin Buzzard}.}
  \bibinfo{year}{2022}\natexlab{}.
\newblock \bibinfo{title}{The Xena Project}.
\newblock \bibinfo{howpublished}{\url{https://xenaproject.wordpress.com/}}.
\newblock
\newblock
\shownote{Accessed: 2022-08-29}.


\bibitem[Carter and Monks(2013)]%
        {lurch}
\bibfield{author}{\bibinfo{person}{Nathan Carter} {and}
  \bibinfo{person}{Kenneth Monks}.} \bibinfo{year}{2013}\natexlab{}.
\newblock \showarticletitle{Lurch: a word processor built on {OpenMath} that
  can check mathematical reasoning}.
\newblock


\bibitem[Chen et~al\mbox{.}(2017)]%
        {orc2a}
\bibfield{author}{\bibinfo{person}{Jianting Chen}, \bibinfo{person}{Medha
  Gopalaswamy}, \bibinfo{person}{Prabir Pradhan}, \bibinfo{person}{Sooji Son},
  {and} \bibinfo{person}{Peter-Michael Osera}.}
  \bibinfo{year}{2017}\natexlab{}.
\newblock \showarticletitle{{ORC$^2$A}: A Proof Assistant for Undergraduate
  Education}. In \bibinfo{booktitle}{\emph{ACM SIGCSE Technical Symposium on
  Computer Science Education}}. \bibinfo{publisher}{Association for Computing
  Machinery}, \bibinfo{address}{Seattle, Washington, USA},
  \bibinfo{pages}{757–758}.
\newblock
\showISBNx{9781450346986}
\urldef\tempurl%
\url{https://doi.org/10.1145/3017680.3022466}
\showDOI{\tempurl}


\bibitem[Church(1940)]%
        {church}
\bibfield{author}{\bibinfo{person}{Alonzo Church}.}
  \bibinfo{year}{1940}\natexlab{}.
\newblock \showarticletitle{A Formulation of the Simple Theory of Types}.
\newblock \bibinfo{journal}{\emph{The Journal of Symbolic Logic}}
  \bibinfo{volume}{5}, \bibinfo{number}{2} (\bibinfo{year}{1940}),
  \bibinfo{pages}{56--68}.
\newblock
\showISSN{00224812}


\bibitem[Coq(2004)]%
        {coq}
\bibfield{author}{\bibinfo{person}{Coq}.} \bibinfo{year}{2004}\natexlab{}.
\newblock \bibinfo{title}{The Coq proof assistant reference manual}.
\newblock
\newblock
\urldef\tempurl%
\url{http://coq.inria.fr}
\showURL{%
\tempurl}
\newblock
\shownote{Version 8.0}.


\bibitem[{de Bruijn}(1972)]%
        {debruijn}
\bibfield{author}{\bibinfo{person}{N.G {de Bruijn}}.}
  \bibinfo{year}{1972}\natexlab{}.
\newblock \showarticletitle{Lambda calculus notation with nameless dummies, a
  tool for automatic formula manipulation, with application to the
  Church-Rosser theorem}.
\newblock \bibinfo{journal}{\emph{Indagationes Mathematicae (Proceedings)}}
  \bibinfo{volume}{75}, \bibinfo{number}{5} (\bibinfo{year}{1972}),
  \bibinfo{pages}{381--392}.
\newblock


\bibitem[Deaduction(2022)]%
        {deaduction}
\bibfield{author}{\bibinfo{person}{Deaduction}.}
  \bibinfo{year}{2022}\natexlab{}.
\newblock \bibinfo{title}{The {Deaduction} GitHub organisation}.
\newblock \bibinfo{howpublished}{\url{https://github.com/dEAduction}}.
\newblock
\newblock
\shownote{Accessed: 2022-08-29}.


\bibitem[Eastlund et~al\mbox{.}(2007)]%
        {eastlund2007}
\bibfield{author}{\bibinfo{person}{Carl Eastlund}, \bibinfo{person}{Dale
  Vaillancourt}, {and} \bibinfo{person}{Matthias Felleisen}.}
  \bibinfo{year}{2007}\natexlab{}.
\newblock \showarticletitle{{ACL2} for freshmen: {First} experiences}. In
  \bibinfo{booktitle}{\emph{ACL2 Workshop}}.
\newblock


\bibitem[Edgcomb et~al\mbox{.}(2015)]%
        {edgcomb2015student}
\bibfield{author}{\bibinfo{person}{Alex~Daniel Edgcomb}, \bibinfo{person}{Frank
  Vahid}, \bibinfo{person}{Roman Lysecky}, \bibinfo{person}{Andre Knoesen},
  \bibinfo{person}{Rajeevan Amirtharajah}, {and} \bibinfo{person}{Mary~Lou
  Dorf}.} \bibinfo{year}{2015}\natexlab{}.
\newblock \showarticletitle{Student performance improvement using interactive
  textbooks: A three-university cross-semester analysis}. In
  \bibinfo{booktitle}{\emph{2015 ASEE Annual Conference \& Exposition}}.
  \bibinfo{pages}{26--1423}.
\newblock


\bibitem[Edukera(2022)]%
        {edukera}
\bibfield{author}{\bibinfo{person}{Edukera}.} \bibinfo{year}{2022}\natexlab{}.
\newblock \bibinfo{title}{The {Edukera} Website}.
\newblock \bibinfo{howpublished}{\url{https://edukera.com/}}.
\newblock
\newblock
\shownote{Accessed: 2022-08-29}.


\bibitem[Gentzen(1935)]%
        {gentzen}
\bibfield{author}{\bibinfo{person}{Gerhard Gentzen}.}
  \bibinfo{year}{1935}\natexlab{}.
\newblock \showarticletitle{Untersuchungen {\"u}ber das logische Schlie{\ss}en.
  I}.
\newblock \bibinfo{journal}{\emph{Mathematische Zeitschrift}}
  \bibinfo{volume}{39}, \bibinfo{number}{1} (\bibinfo{year}{1935}),
  \bibinfo{pages}{176--210}.
\newblock


\bibitem[{ghcjs}(2022)]%
        {ghcjs}
\bibfield{author}{\bibinfo{person}{{ghcjs}}.} \bibinfo{year}{2022}\natexlab{}.
\newblock \bibinfo{title}{The {ghcjs} GitHub Project}.
\newblock \bibinfo{howpublished}{\url{https://github.com/ghcjs/ghcjs}}.
\newblock
\newblock
\shownote{Accessed: 2022-08-29}.


\bibitem[Greenberg and Osborn(2019)]%
        {coqpl19}
\bibfield{author}{\bibinfo{person}{Michael Greenberg} {and}
  \bibinfo{person}{Joseph~C. Osborn}.} \bibinfo{year}{2019}\natexlab{}.
\newblock \showarticletitle{Teaching Discrete Mathematics to Early
  Undergraduates with Software Foundations}.
  \bibinfo{howpublished}{\url{https://cs.pomona.edu/~michael/papers/coqpl2019.pdf}}.
  In \bibinfo{booktitle}{\emph{Workshop on Coq for Programming Languages}}.
  \bibinfo{address}{Lisbon, Portugal}.
\newblock


\bibitem[Hallgren(2022)]%
        {alfa}
\bibfield{author}{\bibinfo{person}{Thomas Hallgren}.}
  \bibinfo{year}{2022}\natexlab{}.
\newblock \bibinfo{title}{Alfa Website}.
\newblock
  \bibinfo{howpublished}{\url{https://cth.altocumulus.org/~hallgren/Alfa/index.html}}.
\newblock
\newblock
\shownote{Accessed: 2022-08-29}.


\bibitem[Harrop(1956)]%
        {harrop}
\bibfield{author}{\bibinfo{person}{Ronald Harrop}.}
  \bibinfo{year}{1956}\natexlab{}.
\newblock \showarticletitle{On disjunctions and existential statements in
  intuitionistic systems of logic}.
\newblock \bibinfo{journal}{\emph{Math. Ann.}} \bibinfo{volume}{132},
  \bibinfo{number}{4} (\bibinfo{year}{1956}), \bibinfo{pages}{347--361}.
\newblock


\bibitem[Huet(1975)]%
        {huet}
\bibfield{author}{\bibinfo{person}{G\'erard~P. Huet}.}
  \bibinfo{year}{1975}\natexlab{}.
\newblock \showarticletitle{A unification algorithm for typed
  $\lambda$-calculus}.
\newblock \bibinfo{journal}{\emph{Theoretical Computer Science}}
  \bibinfo{volume}{1}, \bibinfo{number}{1} (\bibinfo{year}{1975}),
  \bibinfo{pages}{27--57}.
\newblock


\bibitem[Klein et~al\mbox{.}(2009)]%
        {Klein:2009}
\bibfield{author}{\bibinfo{person}{Gerwin Klein}, \bibinfo{person}{Kevin
  Elphinstone}, \bibinfo{person}{Gernot Heiser}, \bibinfo{person}{June
  Andronick}, \bibinfo{person}{David Cock}, \bibinfo{person}{Philip Derrin},
  \bibinfo{person}{Dhammika Elkaduwe}, \bibinfo{person}{Kai Engelhardt},
  \bibinfo{person}{Rafal Kolanski}, \bibinfo{person}{Michael Norrish},
  \bibinfo{person}{Thomas Sewell}, \bibinfo{person}{Harvey Tuch}, {and}
  \bibinfo{person}{Simon Winwood}.} \bibinfo{year}{2009}\natexlab{}.
\newblock \showarticletitle{{seL4}: Formal Verification of an {OS} Kernel}. In
  \bibinfo{booktitle}{\emph{ACM Symposium on Operating Systems Principles}}.
  \bibinfo{publisher}{ACM}, \bibinfo{address}{Big Sky, MT, USA},
  \bibinfo{pages}{207--220}.
\newblock


\bibitem[Knobelsdorf et~al\mbox{.}(2017)]%
        {evalstudy}
\bibfield{author}{\bibinfo{person}{Maria Knobelsdorf},
  \bibinfo{person}{Christiane Frede}, \bibinfo{person}{Sebastian B\"{o}hne},
  {and} \bibinfo{person}{Christoph Kreitz}.} \bibinfo{year}{2017}\natexlab{}.
\newblock \showarticletitle{Theorem Provers as a Learning Tool in Theory of
  Computation}. In \bibinfo{booktitle}{\emph{ACM Conference on International
  Computing Education Research}}. \bibinfo{publisher}{Association for Computing
  Machinery}, \bibinfo{address}{Tacoma, Washington, USA},
  \bibinfo{pages}{83–92}.
\newblock
\showISBNx{9781450349680}
\urldef\tempurl%
\url{https://doi.org/10.1145/3105726.3106184}
\showDOI{\tempurl}


\bibitem[Miller(1996)]%
        {miller}
\bibfield{author}{\bibinfo{person}{Dale Miller}.}
  \bibinfo{year}{1996}\natexlab{}.
\newblock \showarticletitle{A Logic Programming Language With
  Lambda-Abstraction, Function Variables, and Simple Unification}.
\newblock \bibinfo{journal}{\emph{Journal of Logic and Computation}}
  \bibinfo{volume}{1} (\bibinfo{date}{07} \bibinfo{year}{1996}).
\newblock
\showISBNx{3-540-53590-X}
\urldef\tempurl%
\url{https://doi.org/10.1093/logcom/1.4.497}
\showDOI{\tempurl}


\bibitem[Narboux(2005)]%
        {narboux}
\bibfield{author}{\bibinfo{person}{Julien Narboux}.}
  \bibinfo{year}{2005}\natexlab{}.
\newblock \showarticletitle{Toward the use of a proof assistant to teach
  mathematics}.
\newblock \bibinfo{journal}{\emph{International Conference on Technology in
  Mathematics Teaching}} (\bibinfo{date}{07} \bibinfo{year}{2005}).
\newblock


\bibitem[Nipkow(1993)]%
        {nipkow}
\bibfield{author}{\bibinfo{person}{Tobias Nipkow}.}
  \bibinfo{year}{1993}\natexlab{}.
\newblock \showarticletitle{Functional Unification of Higher-Order Patterns}.
  In \bibinfo{booktitle}{\emph{Proceedings of the 8th IEEE Symposium on Logic
  in Computer Science}}. \bibinfo{pages}{64--74}.
\newblock


\bibitem[Nipkow and Klein(2014)]%
        {nipkowklein}
\bibfield{author}{\bibinfo{person}{Tobias Nipkow} {and} \bibinfo{person}{Gerwin
  Klein}.} \bibinfo{year}{2014}\natexlab{}.
\newblock \bibinfo{booktitle}{\emph{Concrete Semantics: With Isabelle/HOL}}.
\newblock \bibinfo{publisher}{Springer Publishing Company, Incorporated}.
\newblock
\showISBNx{3319105418}


\bibitem[Nipkow et~al\mbox{.}(2002)]%
        {isabelle}
\bibfield{author}{\bibinfo{person}{Tobias Nipkow}, \bibinfo{person}{Lawrence~C
  Paulson}, {and} \bibinfo{person}{Markus Wenzel}.}
  \bibinfo{year}{2002}\natexlab{}.
\newblock \bibinfo{booktitle}{\emph{Isabelle/HOL: a proof assistant for
  higher-order logic}}. Vol.~\bibinfo{volume}{2283}.
\newblock \bibinfo{publisher}{Springer Science \& Business Media}.
\newblock


\bibitem[Norell(2009)]%
        {agda}
\bibfield{author}{\bibinfo{person}{Ulf Norell}.}
  \bibinfo{year}{2009}\natexlab{}.
\newblock \showarticletitle{Dependently Typed Programming in Agda}. In
  \bibinfo{booktitle}{\emph{Proceedings of the 4th International Workshop on
  Types in Language Design and Implementation}} (Savannah, GA, USA)
  \emph{(\bibinfo{series}{TLDI '09})}. \bibinfo{publisher}{Association for
  Computing Machinery}, \bibinfo{address}{New York, NY, USA},
  \bibinfo{pages}{1–2}.
\newblock
\showISBNx{9781605584201}
\urldef\tempurl%
\url{https://doi.org/10.1145/1481861.1481862}
\showDOI{\tempurl}


\bibitem[Pfenning and Sch{\"u}rmann(1999)]%
        {twelf}
\bibfield{author}{\bibinfo{person}{Frank Pfenning} {and}
  \bibinfo{person}{Carsten Sch{\"u}rmann}.} \bibinfo{year}{1999}\natexlab{}.
\newblock \showarticletitle{System Description: {Twelf} --- A Meta-Logical
  Framework for Deductive Systems}. In \bibinfo{booktitle}{\emph{Automated
  Deduction --- CADE-16}}. \bibinfo{publisher}{Springer Berlin Heidelberg},
  \bibinfo{address}{Berlin, Heidelberg}, \bibinfo{pages}{202--206}.
\newblock


\bibitem[Pierce(2009)]%
        {pierce}
\bibfield{author}{\bibinfo{person}{Benjamin~C. Pierce}.}
  \bibinfo{year}{2009}\natexlab{}.
\newblock \showarticletitle{Lambda, the Ultimate {TA}: {Using} a Proof
  Assistant to Teach Programming Language Foundations}. In
  \bibinfo{booktitle}{\emph{International Conference on Functional
  Programming}}. \bibinfo{publisher}{Association for Computing Machinery},
  \bibinfo{address}{Edinburgh, Scotland}, \bibinfo{pages}{121--122}.
\newblock
\showISBNx{9781605583327}
\urldef\tempurl%
\url{https://doi.org/10.1145/1596550.1596552}
\showDOI{\tempurl}


\bibitem[Pierce et~al\mbox{.}(2017)]%
        {software_foundations}
\bibfield{author}{\bibinfo{person}{Benjamin~C. Pierce}, \bibinfo{person}{Arthur
  {Azevedo de Amorim}}, \bibinfo{person}{Chris Casinghino},
  \bibinfo{person}{Marco Gaboardi}, \bibinfo{person}{Michael Greenberg},
  \bibinfo{person}{C\v{a}t\v{a}lin Hri\c{t}cu}, \bibinfo{person}{Vilhelm
  Sj\"{o}berg}, {and} \bibinfo{person}{Brent Yorgey}.}
  \bibinfo{year}{2017}\natexlab{}.
\newblock \bibinfo{booktitle}{\emph{Software Foundations}}.
\newblock \bibinfo{publisher}{Electronic textbook}.
\newblock
\urldef\tempurl%
\url{http://www.cis.upenn.edu/~bcpierce/sf}
\showURL{%
\tempurl}
\newblock
\shownote{Version 6.1}.


\bibitem[Robinson(1965)]%
        {robinson}
\bibfield{author}{\bibinfo{person}{John~A. Robinson}.}
  \bibinfo{year}{1965}\natexlab{}.
\newblock \showarticletitle{A Machine-Oriented Logic Based on the Resolution
  Principle}.
\newblock \bibinfo{journal}{\emph{J. ACM}} \bibinfo{volume}{12},
  \bibinfo{number}{1} (\bibinfo{date}{jan} \bibinfo{year}{1965}),
  \bibinfo{pages}{23--41}.
\newblock
\showISSN{0004-5411}
\urldef\tempurl%
\url{https://doi.org/10.1145/321250.321253}
\showDOI{\tempurl}


\bibitem[Wadler et~al\mbox{.}(2022)]%
        {plfa}
\bibfield{author}{\bibinfo{person}{Philip Wadler}, \bibinfo{person}{Wen Kokke},
  {and} \bibinfo{person}{Jeremy~G. Siek}.} \bibinfo{year}{2022}\natexlab{}.
\newblock \bibinfo{booktitle}{\emph{Programming Language Foundations in
  {A}gda}}.
\newblock
\urldef\tempurl%
\url{https://plfa.inf.ed.ac.uk/20.08/}
\showURL{%
\tempurl}


\bibitem[Yang(2022a)]%
        {interactivesequent}
\bibfield{author}{\bibinfo{person}{Edward~Z. Yang}.}
  \bibinfo{year}{2022}\natexlab{a}.
\newblock \bibinfo{title}{Interactive Tutorial of the Sequent Calculus}.
\newblock
  \bibinfo{howpublished}{\url{http://logitext.mit.edu/logitext.fcgi/tutorial}}.
\newblock
\newblock
\shownote{Accessed: 2022-08-29}.


\bibitem[Yang(2022b)]%
        {logitext}
\bibfield{author}{\bibinfo{person}{Edward~Z. Yang}.}
  \bibinfo{year}{2022}\natexlab{b}.
\newblock \bibinfo{title}{Logitext}.
\newblock \bibinfo{howpublished}{\url{http://logitext.mit.edu/main}}.
\newblock
\newblock
\shownote{Accessed: 2022-08-29}.


\bibitem[Zabarauskas(2022)]%
        {zabr}
\bibfield{author}{\bibinfo{person}{Brendan Zabarauskas}.}
  \bibinfo{year}{2022}\natexlab{}.
\newblock \bibinfo{title}{{Martin-L\"of} Type Theory in {Holbert}}.
\newblock
  \bibinfo{howpublished}{\url{https://gist.github.com/brendanzab/1b4732179b15201bf33fed6dbca02458}}.
\newblock
\newblock
\shownote{Accessed: 2022-08-29}.


\end{thebibliography}

\end{document}